\documentclass[fleqn,11p,times]{elsarticle}
\usepackage[utf8]{inputenc}
\usepackage{setspace}
\usepackage{graphicx}
\usepackage{graphics}
\usepackage{xcolor}
\definecolor{navy}{rgb}{0.000000,0.000000,0.501961}
\graphicspath{{Images/}}
\usepackage[labelfont=bf]{caption}
\usepackage{subcaption}

\usepackage{setspace}

\usepackage[ruled,linesnumbered]{algorithm2e}

\usepackage{amsmath}

\usepackage{amsthm}
\makeatletter
\def\els@aparagraph[#1]#2{\elsparagraph[#1]{#2\@addpunct{.}}}
\def\els@bparagraph#1{\elsparagraph*{#1\@addpunct{.}}}
\SetAlgoCaptionSeparator{.}
\RestyleAlgo{algoruled}

\makeatother

\usepackage{float}
\usepackage{url}



\begin{document}

\begin{frontmatter}

\title{Performance assessment of vehicle interdiction strategies in a simulation based environment on a complex transportation network}

\vspace{0.4cm}
\author[1]{Sukanya Samanta\corref{cor1}}
\ead{ssukanya@iitkgp.ac.in }
\cortext[cor1]{Corresponding author}

\author[2]{Jatin Uniyal}

\author[3]{Goutam Sen}

\author[4]{Soumya Kanti Ghosh}

\vspace{0.4cm}

\address[1]{Advanced Technology Development Centre, Indian Institute of Technology Kharagpur, Kharagpur, West Bengal 721302, India}
\address[2]{Department of Industrial and Systems Engineering, Indian Institute of Technology Kharagpur, Kharagpur, West Bengal 721302, India}
\address[3]{Department of Industrial and Systems Engineering, Indian Institute of Technology Kharagpur, Kharagpur, West Bengal 721302, India}
\address[4]{Department of Computer Science and Engineering, Indian Institute of Technology Kharagpur, Kharagpur, West Bengal 721302, India}

\vspace{0.4cm}

\begin{abstract}
We consider the escape interdiction problem in a transportation network. In the absence of traffic in the network, the criminal/attacker tries to escape from the city using any of the shortest paths from the crime scene to any randomly chosen exit point. In the presence of traffic, the attacker chooses the optimal path, which takes minimum time to reach his destination. On the contrary, police/defenders try to interdict the criminal on his escape route. This is a challenging task for police with limited resources. Again, a real city road network is also complex in nature. First, we develop a simulation-based model for the optimal allocation of resources using the SUMO simulator. Next, we focus on developing a more advanced search strategy like routing with optimal resource allocation. We develop a novel meta-heuristic approach in a simulation environment to interdict the attacker in a dynamic crime scenario. Like the previous approach, the attacker follows the path with optimal time to escape from the city. In contrast, defenders try to catch the attacker regardless of the path which the attacker takes. The defenders aim is to maximize the interdiction probability. As time plays a vital role, we choose a meta-heuristic approach to provide quality solutions in a time-efficient manner. We test the developed meta-heuristic with a case study on the IIT Kharagpur map, India. We analyze the performance of the mentioned approaches using the SUMO simulator both in the presence of traffic and without traffic. We develop a linear regression model to generate optimal path in the presence of traffic. Here traffic is generated randomly in the whole network using the SUMO simulator.
\end{abstract}
\begin{keyword}
\texttt{Resource allocation, Genetic algorithm, Vehicle routing and scheduling, Network optimization, Linear regression model, SUMO simulator}
\end{keyword}
\end{frontmatter}

\section{Introduction}
\label{S:1}
Police patrol and police resource allocation are vital components in crime reduction strategies of the law enforcement agencies (Samanta et al. \cite{samanta2021literature}). The limited resources like the personnel and the vehicles must be deployed optimally. This research aims to maximize the interdiction rate considering a moving attacker vehicle in a given urban area road network. Hence, the situation finds the dynamic crime scene where the location of the attacker vehicle is changing over time. We consider one criminal vehicle and multiple police vehicles. In this case, we consider assumption like defenders know the initial location of the crime. For example, in the case of a bank robbery, the bank's location is known to defenders. Because of the nature of these crimes, defenders have time to react and catch the criminal before he escapes from the city. Hence, a strategy that only addresses the static crime is insufficient in this case, which is the novelty of this problem. As the city road networks are complex and defenders resources are also limited in number, decision-making in every step has a significant role in this dynamic scenario. Hence, search strategies for defenders need to be developed to interdict the attacker vehicle.

First, we consider the problem of static defender vehicle allocation (SDVA) in complex road networks. The problem considers the static allocation of police resources at the most critical junctions in a large transportation network. Therefore, an effective resource allocation strategy for police vehicles needs to be developed to catch the criminal. Samanta et al. \cite{samanta2021vehicle} propose an optimal resource allocation approach in a simulation-based environment as it is hard to model the problem mathematically. They test the developed approach with a case study on the IIT Kharagpur road network, India. Using the SUMO simulator, they simulate the total simulation time, the number of vehicles, vehicle speed but traffic and traffic signals are not considered in the developed model. Hence, in this simulation work, traffic is generated with the help of \textit{duarouter} application of SUMO simulator. In addition, the traffic signal at the selected junction is generated using the \textit{netedit} application of the SUMO simulator. In case of the simulation model with traffic, we show that the interdiction probability increases with the increase of traffic, which is expected due to the availability of larger search time window.

Second, we introduce the defender vehicle routing and scheduling problem (DVRSP), in which the police vehicles make a tour in the network instead of a single halt. So, the network search options increase exponentially, and optimization is required to identify the best routes and schedules for the vehicles. Interestingly, traditional multi-depot vehicle routing problem (VRP) might seem analogous, but there are some fundamental differences in the problem features and constraints. In this case, the visit points are not known in advance. Instead, a set of potential nodes on the shortest path of the attacker is given. The vehicles need to choose the points to generate a tour. However, all points may not be covered. Secondly, the vehicles do not return to the depot and go on searching in the network till the maximum allowable time. Thirdly, the same node can be visited by two different police vehicles at different points of time. So, the visits are not for customer demand satisfaction but rather to check whether the attacker passes through that node or not. 

In case of both approaches, we assume that the attacker tries to reach the randomly chosen exit point from the crime scene using any of the available shortest paths. When there is no traffic in the network, Dijkstra's algorithm generates the optimal path, which is the shortest path. But in the presence of traffic in the network, the optimal path may not always be the shortest path because of possible traffic congestion. In the presence of traffic, the optimal path refers to the path which takes minimum time to reach from source to destination. The optimal path is determined with the help of a linear regression model, which is used alongside Dijkstra's algorithm. To develop this regression model, a synthetic dataset is generated using the SUMO simulator. 

The challenge lies in developing the strategy for defenders, which is a collection of nodes along with the start and end time attached to each node. The MILP based formulations face time complexity and scalability issues. As solution time plays a vital role in this type of exigency, we choose GA to provide good quality solutions quickly. In addition, the developed GA scales well to large road networks. The output of GA is fed into the SUMO simulator and we checked the performance without traffic as well as in presence of traffic. SUMO is a microscopic traffic simulator and is also used for traffic light scheduling (Dias et al. \cite{dias2014inverted}, Garc{\'\i}a-Nieto et al. \cite{garcia2012swarm}).

We can summarize the main contributions of this paper as follows. (1) This is the first attempt to develop a comparison-based study in a simulation environment for escape interdiction problems. The focus is given on the development of solution methodologies. (2) We develop an optimal allocation approach of resources in the presence of traffic, and we are not aware of any such implementation earlier for the escape interdiction problems. (3) In addition, we develop a novel GA-based meta-heuristic approach in a simulation environment. In the presence of traffic, we generate optimal paths for defenders where each node is associated with a fixed time window which is the waiting time for the defender. (4) We also develop a linear regression model to find out the optimal path in the presence of traffic.

The paper is organized as follows. In Section 2, we present the relevant research. In Sections 3 and 4, we define the proposed approaches for the mentioned SDVA and DVRSP problems, respectively. Section 5 presents the synthetic traffic dataset generation and the development of a linear regression model. We present the quality of the results in Section 6. Section 6 also documents a case study on the IIT Kharagpur map, India. We conclude this research in Section 7.

\section{Related work}
\label{S:2}

To interdict attackers efficiently, optimal resource allocation and route design for defenders are challenging with limited available resources. Many papers focus on the optimal allocation of resources. For example, Shang et al. \cite{shang2018operation} consider an optimal allocation of defender resources, and Bier et al.\cite{bier2008optimal} focus on the cost-effective resource allocation among the potential terrorist targets. They develop a rigorous computational model to allocate resources considering the uncertainty about the attacker’s target locations. Similarly, Paruchuri et al. \cite{paruchuri2007efficient} develop a heuristic named ASAP to find out the highest-reward strategy for defenders considering security against multiple adversaries. Again, Tsai et al. \cite{tsai2009iris} focus on the efficient resource allocation in transportation networks. They develop the Intelligent Randomization In Scheduling (IRIS) system which is a software scheduling assistant for the Federal Air Marshals (FAMs). Similarly, Kiekintveld et al. \cite{kiekintveld2009computing} develop new algorithms for optimal randomized resource allocation with the increasing problem size at the Los Angeles International Airport (LAX). Several papers focus on the improvement of resource allocation strategy considering security games Tsai et al. \cite{tsai2010urban}, Yang et al. \cite{yang2011improving}, and Yang et al. \cite{yang2012modeling}. Whereas Cappanera and Scaparra \cite{cappanera2011optimal} develop a multilevel optimization model and an enumeration algorithm for cost-effective resource allocation in shortest-path networks. Again, Fang et al. \cite{fang2013designing} and Fang et al. \cite{fang2013protecting} focus on scheduling security resources to protect moving targets. Hence, optimal resource allocation and route design can reduce crime significantly. 

Samanta et al. \cite{samanta2022vns} develop variable neighborhood search based meta-heuristic approach for both defenders and attacker to interdict the attacker optimally in a dynamic crime scenario. In this case, near-optimal defender strategies and attacker strategy are generated once the Nash Equilibrium is reached. This developed approach is not suitable in a simulated environment as in each iteration the entire simulation environment needs to be created which will be more time consuming for a large transport network. In addition, traffic and traffic lights are not considered in the developed model. Again, playing multiple strategies with mixed probabilities is impossible in practical scenario as the mixed strategies cannot be fed to the vehicles in a simulation based environment. Hence we focus on developing GA based meta-heuristic approach to generated near-optimal search strategies for defenders in a simulation environment.

The GA is a well-known optimizer which can provide near-optimal solutions quickly. Again, the operators like selection, crossover, and mutation make GA an excellent approach to solve complex real-world problems. Using these operators of GA, exploration and exploitation can be achieved easily. Hence, we propose a GA based meta-heuristic for defenders to generate near-optimal solutions in a time-efficient manner for large road networks. Solution quality can also be checked by defining a proper fitness function in GA (Whitley \cite{whitley1994genetic}, Harik et al. \cite{harik1999compact}). Again, Thangiah \cite{thangiah1993vehicle} develops a GA based methodology for vehicle routing problem with time windows and tries to minimize service cost with limited vehicle capacity and travel time because time always plays an important role in vehicle routing problems. Similarly, in this paper, the objective is to maximize the rate with which defenders can catch the attacker in a large transportation network.

\section{Proposed SDVA approach}
\label{S:3}
This section presents the proposed methodology for the considered SDVA problem. In SDVA, we adopt similar approach for resource allocation which is developed by Samanta et al. \cite{samanta2021vehicle}. Here the checkpoints are chosen depending on the betweenness centrality measure but we optimally allocate the police resources to those checkpoints using the Hungarian algorithm. The betweenness centrality of a node is the number of the shortest paths that pass through that node, considering each pair of nodes in that graph. The betweenness centrality of a node $u$ is represented by $b(u)$ where $b(u) = \sum\limits_{m\neq u \neq n} \sigma_{mn} (u) / \sigma_{mn}$. Here, $\sigma_{mn}$ is the total number of shortest paths from $m$ to $n$ and $\sigma_{mn} (u)$ is the number of those paths that pass through $u$ considering $u$ is not an end point. Attacker takes the shortest path from the random crime location to the randomly selected exit point. Considering the without traffic scenario, the path cost is calculated depending on the path length using Dijkstra's algorithm. The Hungarian algorithm is applied to the cost matrix which is generated depending on the path cost from each initial location of police to each halting point. The developed SDVA approach is shown in Algorithm~\ref{alg:30}. Here $max_{iterations}$ is the maximum number of iterations before the simulation ends. 

\begin{figure}[H]
\centering
\includegraphics[scale=0.5]{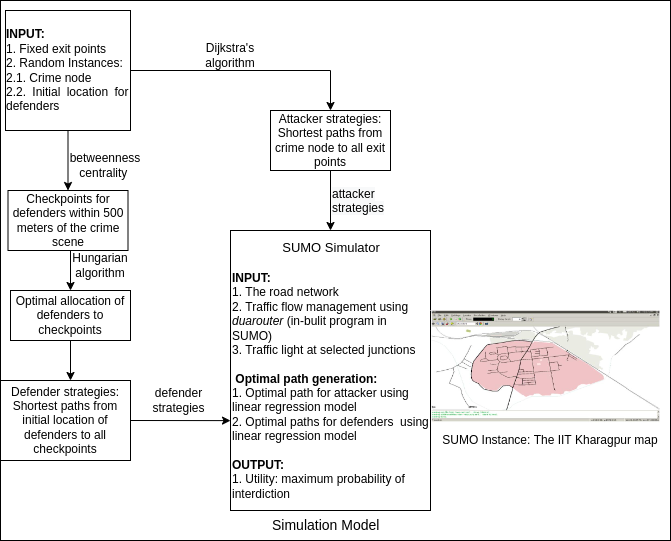}
\caption{Optimal allocation of defender vehicles in a simulation environment with traffic}
\label{image-f75}
\end{figure}

In the presence of traffic, the optimal path is not simply the shortest path. The optimal path depends on the traffic conditions in the network. So, the linear regression model is used to determine the optimal path and the corresponding travel cost in the presence of traffic in the transportation network. Then, like the without traffic scenario, we follow the same previous steps of Algorithm~\ref{alg:30} to evaluate the performance of the developed simulation model for SDVA with a traffic scenario (see Fig. \ref{image-f75}).

\begin{algorithm}[H]
\SetAlgoLined
 \textbf{Input:} Fixed exit points, Betweenness measures for all junctions;\\
 \textbf{Output:} Utility;\\
  \While{$iterations < max_{iterations}$}{
    \textbf{bestAttackerPathGeneration (Fixed exit points)}: \newline
        \textbf{4.1} $\;$ Generate random crime scene; \newline
        \textbf{4.2} $\;$ Select one exit point randomly; \newline
        \textbf{4.3} $\;$ Calculate the shortest path from the crime scene to the randomly selected exit point using Dijkstra’s algorithm for the attacker\;
    \textbf{bestDefendersPathGeneration (betweenness measures)}: \newline
        \textbf{5.1} $\;$ Generate random initial locations for defenders; \newline
        \textbf{5.2} $\;$ Select the critical junctions for defenders using betweenness centrality measure and also the critical junctions must be within $distance_{min}$ from crime scene; \newline
        \textbf{5.3} $\;$Optimal allocation of defender vehicles to those critical junctions using Hungarian algorithm; \newline
         \textbf{5.4} $\;$ Generate path for defenders from initial locations to those critical junctions using Dijkstra’s algorithm\;
    \textbf{runSimulation}: \newline
         \textbf{6.1} $\;$ Allocate the generated path to the attacker; \newline
        \textbf{6.2} $\;$ Allocate the generated paths to the defenders; \newline
       \textbf{6.3} $\;$ \textbf{If}\textit{ distance between the criminal vehicle and any defender vehicle $< d_{min}$ meters }\textbf{then} 
            Success = Success + 1;\newline  
        \textbf{6.4} $\;$ \textbf{If}\textit{ distance between the criminal vehicle and exit point $< d_{min}$ }\textbf{then} 
            Failure\;
   }
   Calculate the final Utility depending on the Success rate\;
   \Return (Utility).
  \caption{Static Defender Vehicle Allocation in a Simulation Environment (SDVASE)}
 \label{alg:30}
 \end{algorithm}

\begin{figure}[H]
\centering
\includegraphics[scale=0.44]{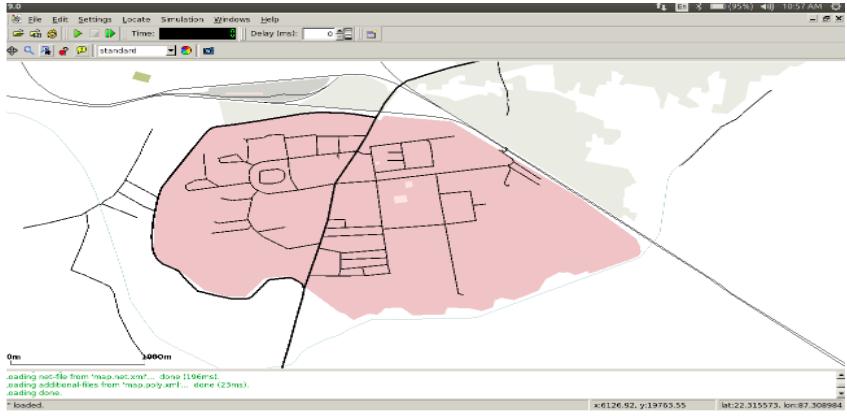}
\caption{IIT Kharagpur map in SUMO simulator}
\label{image-f71}
\end{figure}

In Fig. \ref{image-f71}, the IIT Kharagpur map is imported into the Sumo simulator from the OpenStreetMap to implement the proposed SDVA approach. We use the in-built TraCI functions like $\textit{traci.vehicle.setStop(vehID, edgeID), traci.vehicle.add(vehID, routeID)}$ etc to implement the simulation model.

\begin{figure}[H]
\centering
\includegraphics[scale=0.2]{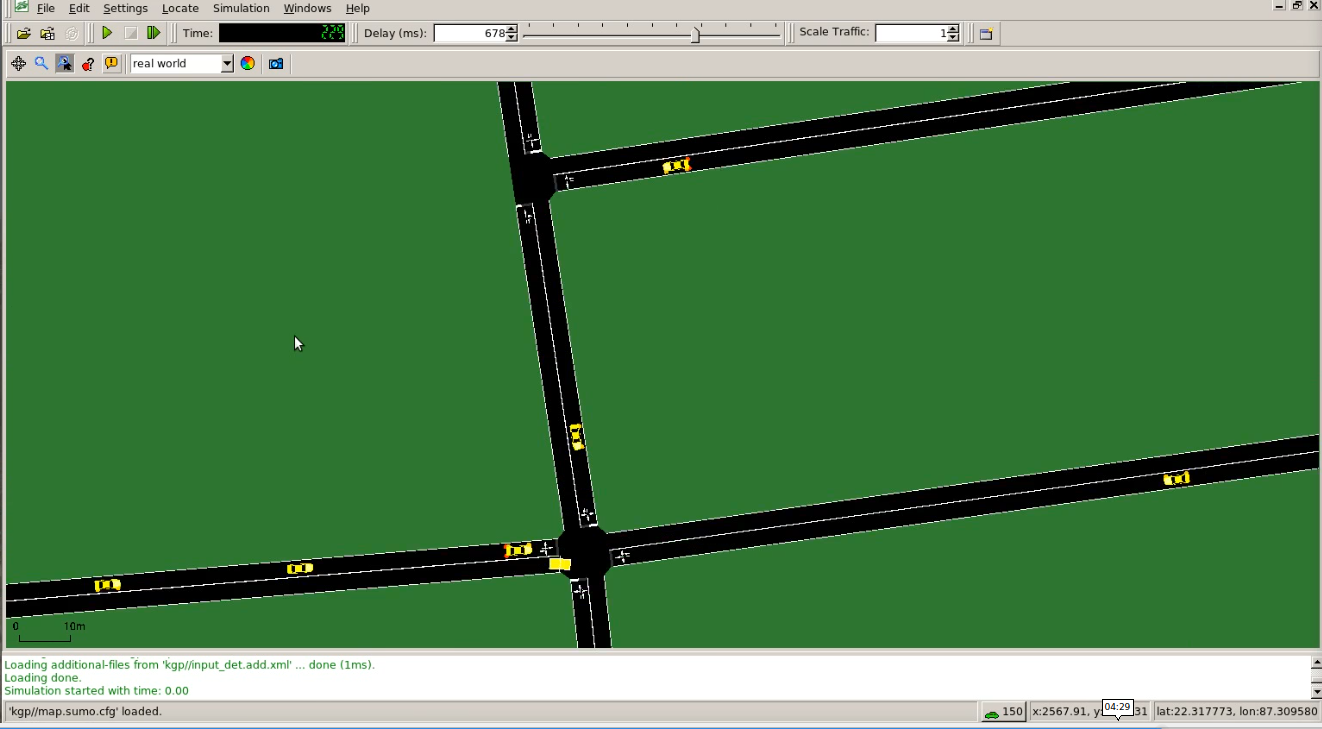}
\caption{Traffic on the IIT Kharagpur map}
\label{image-ft3}
\end{figure}

\begin{figure}[H]
\centering
\includegraphics[scale=0.2]{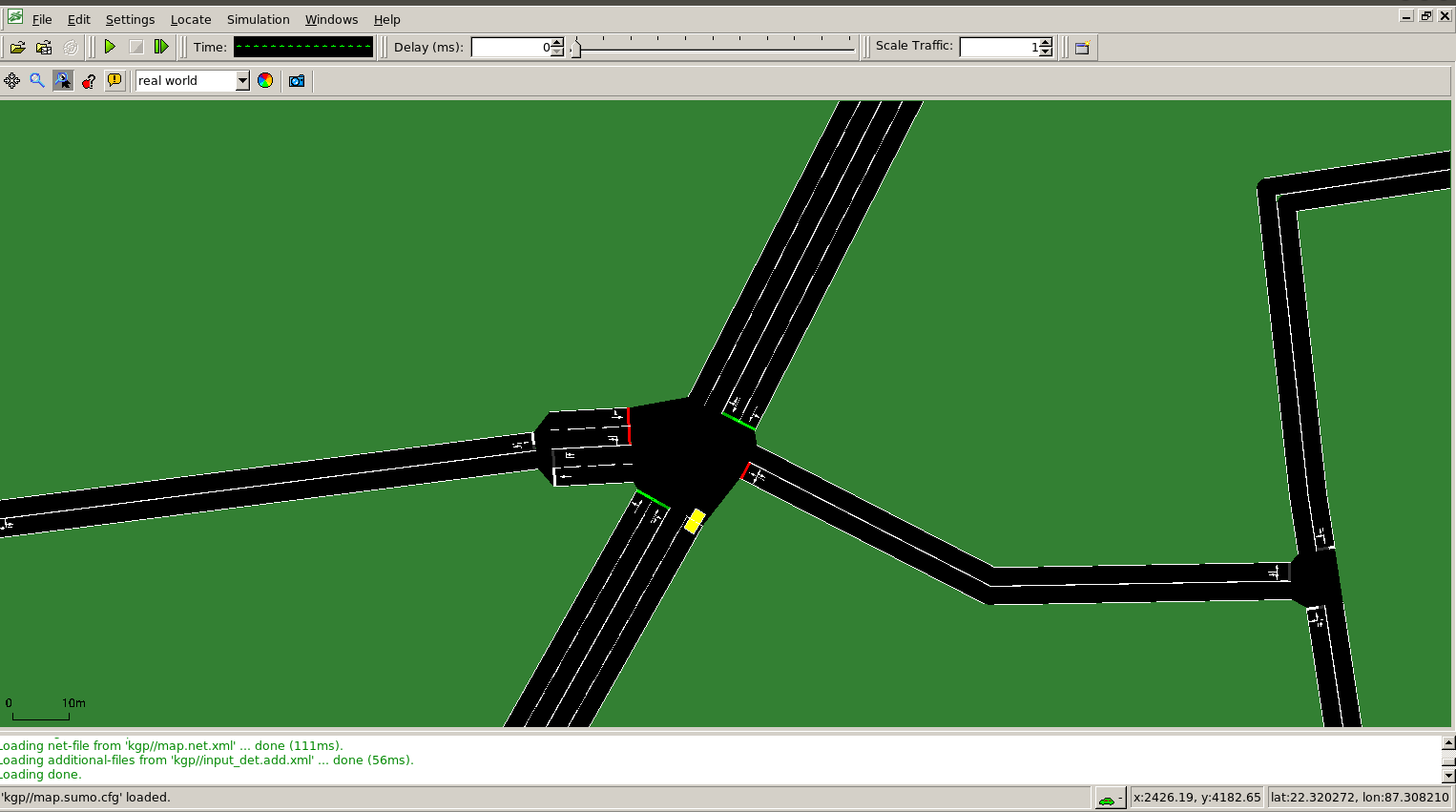}
\caption{Traffic signal scenario on the IIT Kharagpur map}
\label{image-fs1}
\end{figure}

In presence of traffic, we incorporate the effect of traffic density (see Fig. \ref{image-ft3}). In addition, we include traffic signals at the selected junction using the \textit{netedit} application of the SUMO simulator (see Fig. \ref{image-fs1}).

\section{Proposed DVRSP approach}
\label{S:4}

In case of DVRSP, we use similar notations of Zhang et al. \cite{zhang2017optimal} to design our problem. In addition, we adopt the MILP model for defenders which is developed by Zhang et al. \cite{zhang2017optimal} to compare our result with. This MILP model for defenders named bestDo generates the optimal defender strategies but faces time complexity issue for large transportation networks. Hence we focus on developing efficient solutions for complex road networks. We consider that the attacker tries to escape from the crime scene using any of the exit nodes. Here, the attacker takes any of the shortest paths available from the crime location to any of the exit points which is chosen randomly. The defenders traverse to various nodes and halt for specific time intervals to interdict the attacker.  

In this approach, the main focus is given on analysis and decision-making of the optimal strategy for the defenders. Here, the decision space is constructed by considering all permutations of routes (that can be traversed in the graph) and a continuous time domain. Thus the decision space is quite large. For example, in case of IIT Kharagpur map, the number of possible combinations of routes is also quite large. So, the focus is to generate the optimal solution efficiently.

\begin{figure}[H]
\centering
\includegraphics[scale=0.44]{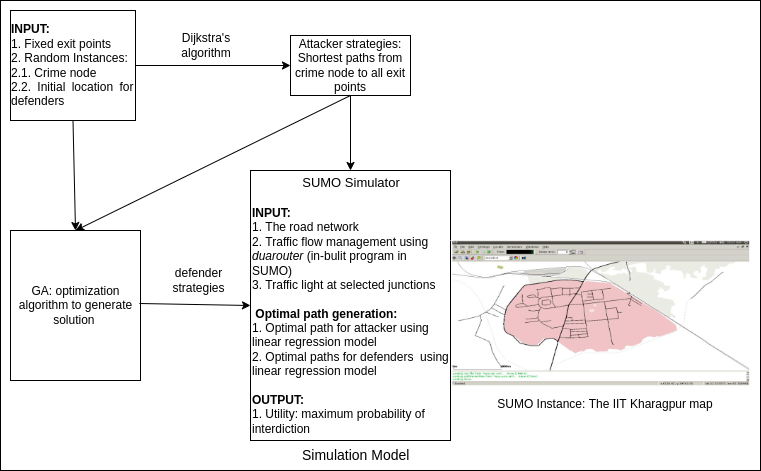}
\caption{Running GA in a simulation environment with traffic}
\label{image-f72}
\end{figure}

In case of DVRSP, the strategies for the defender vehicles are obtained using the developed GA-based meta-heuristic. Again, in the absence of traffic, Dijkstra's algorithm can be used to calculate the optimal path between any two nodes based on the shortest distance only. While in the presence of traffic, we use the developed linear regression model to select the optimal path. Finally, we run the simulations to calculate the probability of interdiction (see Fig. \ref{image-f72}).

In the developed meta-heuristic approach, the defender's strategy is a sequence of states starting from the initial position of the defender and also shows the defender's movements. Since the solution presentation of the problem has a sequential structure, popular heuristic techniques like evolutionary algorithms, simulated annealing, and particle swarm optimization techniques can be used. We choose meta-heuristic because popular literature on this subject uses meta-heuristic to address scalability because of the larger flexibility in the operators and the ease of customization with respect to the problem techniques. Again, meta-heuristic works on random search techniques, and problem based intelligence can be incorporated at various stages of the algorithm to drive it to a near-optimal solution.

\begin{figure}[H]
\centering
\includegraphics[scale=0.55]{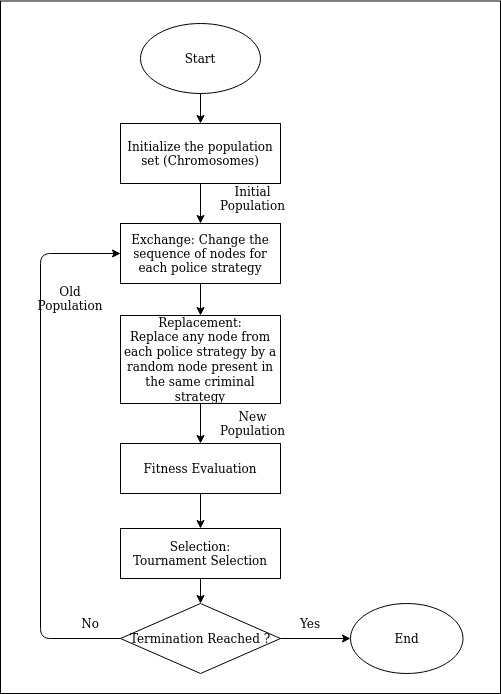}
\caption{Flowchart of the GA}
\label{image-im34}
\end{figure}

Here, each chromosome is presented using an array of vertices for all defenders. In exchange, the positions of nodes are exchanged for each defender in a chromosome. In replacement, one node is picked up randomly from a chromosome and is replaced by another node from the same attacker strategy from which the earlier node belongs. The fitness value of a chromosome is evaluated by its probability of interdiction with respect to the attacker strategies in consideration. Based on the fitness value, tournament selection is done to select the new set of chromosomes, which indicates the new population for the exchange. The best chromosome is selected after $K_{max}$ iterations, which is considered as the termination condition. The flowchart of the proposed meta-heuristic is shown in Fig. \ref{image-im34}.

\subsection{Input: attacker strategies}
\label{S:4.1}

We provide the attacker strategies as input depending on the number of fixed exit points. For example, at least five attacker strategies must be given as input to the heuristic algorithm for five fixed exit points. Here, each attacker strategies has a different exit node. In attacker strategy, each state is represented by $a_j  = (v_j ,  t^a_j )$, which indicates that the attacker reaches at node $v_j$ at time $t^a_j$.

\subsection{Chromosome presentation}
\label{S:4.2}

Initially, we consider ten chromosomes in the population set. Each defender contributes one strategy to form one chromosome. If there are five attacker strategies, each defender's strategy contains five nodes, where each node belongs to a different attacker strategy (see Algorithm~\ref{alg:31}). So, if we consider four defenders, each chromosome consists of a total of twenty nodes.

\begin{algorithm}[H]
\SetAlgoLined
 \textbf{Input:} Initialize attacker strategy set A\\
 \textbf{Output:} Chromosome \\
 \For{all defenders}{
   \For{all attacker strategies}{
        Choose one node to form chromosome
   }
   }
   \Return (chromosome).
  \caption{Chromosome presentation}
 \label{alg:31}
 \end{algorithm}

\subsection{Exchange}
\label{S:4.3}

Exchange is done to change the sequence of nodes for each police as police visit the nodes according to the sequence of the nodes in the strategy (see Algorithm~\ref{alg:32}).

\begin{algorithm}[H]
\SetAlgoLined
 \textbf{Input:} Initialize population set\\
 \textbf{Output:} Chromosome \\
 \For{all chromosome in population set}{
   \For{all defenders}{
        Change the position of nodes in random order
   }
   }
   \Return (chromosome).
  \caption{Exchange}
 \label{alg:32}
 \end{algorithm}

\subsection{Replacement}
\label{S:4.4}

Replacement is performed on one node for each defender. If we consider four defenders, replacement is performed four times for each chromosome. In the replacement operation, the selected node for a particular defender is replaced by any other node from the same attacker strategy. Now, the strategy number is retrieved by the node’s index value. The pseudocode for replacement is shown in Algorithm~\ref{alg:33}.

\begin{algorithm}[H]
\SetAlgoLined
 \textbf{Input:} Initialize population set\\
 \textbf{Output:} Chromosome \\
 \For{all chromosome in population set}{
   \For{all defenders}{
        Choose one node to be replaced by a different node from the same attacker strategy
   }
   }
   \Return (chromosome).
  \caption{Replacement}
 \label{alg:33}
 \end{algorithm}

\subsection{Evaluation of fitness function}
\label{S:4.5}

The $r^{th}$ defender's state $d_r$ is represented by a tuple $(v^r, t^{r,in}, t^{r,out})$, which indicates that the node $v^r$ is monitored by the defender $d_r$ during $[t^{r,in}, t^{r,out}]$. $sp_{d_r}$ is the speed of the defender $d_r$. The $j^{th}$ state of attacker is represented by a tuple $(v^j, t^a_j)$.

For each chromosome, the strategies of each defender are separated to calculate the fitness value. Then, $[t^{r,in}$ and $t^{r,out}]$ arrays are constructed for each strategy of the defender. Here, $t^{r,in}_{i+1} - t^{r,out}_i = dist(v^r_i , v^r_{i+1}) / sp_{d_r}$. We consider that defenders are at their initial position at time zero.

Considering each defender strategy, if any node matches with any node present in any of the attacker strategy and $t^{in}$ value of the defender node is less than $t^a_j$ value of the attacker node then $t^{out}$ value of the defender node is equal to the $t^a_j$ value. Otherwise, $t^{out}$ value of the defender node is equal to the $t^{in}$.  For $a_j  = (v_j ,  t^a_j)$ and $s^r_i= (v^r_i , t^{r,in}_i , t^{r,out}_i)$, if $v^r_i = v_j$  and $t^{r,in}_i \leq t^a_j$ then $t^{r,out}_i$  =  $t^a_j$  else $t^{r,out}_i$  =  $t^{r,in}_i$.

The attacker strategies which are interdicted by any defender of a particular chromosome, are used to calculate the fitness value of that chromosome. Here, the fitness value of the chromosome is the addition of the mixed probability values of the interdicted attacker strategies. Each attacker strategy is assigned with a probability value known as mixed probability value which indicates the probability of interdiction which defender gets when that particular attacker strategy is interdicted. The sum of those mixed probabilities is one. The fitness calculation is shown in Algorithm~\ref{alg:34}.

\begin{algorithm}[H]
\SetAlgoLined
 \textbf{Input:} Initialize $\textit{A},\textit{S}$\\
 \textbf{Output:} Chromosome \\
 \For{each chromosome}{
   \For{all attacker strategies}{
      \If{$v^r_i = v_j$  and $t^{r,in}_i \leq t^a_j \leq t^{r,out}_i$ \, where $a_j  = (v_j ,  t^a_j)$ and $s^r_i= (v^r_i , t^{r,in}_i , t^{r,out}_i)$}{
   			fitness $\leftarrow$ fitness + $y[i]$ \, where $y[i]$ is the mix probability of the $i^{th}$ attacker strategy\\
   			}  
   }
   }
   \Return (chromosome with the best fitness value).
  \caption{Fitness calculation}
 \label{alg:34}
 \end{algorithm}

\subsection{Selection}
\label{S:4.6}

Tournament selection is used to select the best ten chromosomes from the set of twenty chromosomes based on their fitness values. Hence, input to the tournament selection is the array of fitness values of twenty chromosomes. And, the output consists of the ten best chromosomes which form the new population set.

\subsection{Termination criteria}
\label{S:4.7}

As a termination condition, the total number of iterations is set as $K_{max}$.

\section{Synthetic traffic dataset generation and development of linear regression model and simulations for the performance measure of the developed approaches}
\label{S:5}

In order to compare the performances of the two mentioned approaches, we run simulations using the SUMO simulator. Considering both approaches, defenders travel from source to destination as quickly as possible, taking the optimal path. Now Dijkstra's algorithm can give the optimal path without traffic in the road network. But in the presence of traffic, we need another approach to generate an optimal path to travel from source to destination as quickly as possible. Hence we develop a linear regression model to generate the optimal path between any two given nodes in a complex road network with traffic. The model estimates the travel time along any edge, depending on the current traffic conditions in the network. We assume that traffic on a particular edge depends on the number of vehicles and the speed of the slowest vehicle on that edge. So overall, the travel time on a particular edge depends on the traffic on that edge along with the length of that edge. To develop the model, we consider three independent variables, which are the length of the edge, the number of vehicles on the edge, and the speed of the slowest vehicle on the edge. Here, the travel time on the edge is considered as the dependent variable. 

We generate a synthetic traffic dataset using the SUMO simulator as no real traffic data is available for the IIT Kharagpur network. This synthetic traffic dataset is used to develop the regression model. The dataset consists of 292 data points. It is further split into five separate cases for better accuracy. The dataset contains data about (i) the length of a particular edge, (ii) the number of vehicles on that edge, (iii) the speed of the slowest vehicle on that edge, and (iv) the travel time of the test vehicle on that edge. As mentioned earlier, the first three are the independent variables, while the fourth one is the dependent variable. In order to decide the degrees of the three independent variables, we check the performance of the regression model with different combinations of degrees (deg 1, deg 2, and deg 3) w.r.t. the three independent variables. The obtained adjusted \textit{R}-squared values are shown in Table \ref{tab:table3.77}.

\begin{table}[H]
  \begin{center}
   \caption{\\ Adjusted R-squared values of regression model with different degrees w.r.t. the independent variables}
    \label{tab:table3.77}
    \scalebox{0.6}{
    \begin{tabular}{rrrrr}
    \hline
       Test case & $x_i $(length (m)) & $y_i $(no. of vehicles) & $z_i $(speed of slowest vehicle (m/s)) & \textbf{Adjusted R-squared} \\\hline
       1 & deg 1 &	deg 1 &	deg 1 &	0.8854 \\ \hline
       2 & deg 1 &	deg 1 &	deg 2 &	0.8993 \\ \hline
       3 & deg 1 &	deg 1 &	deg 3 &	0.8985 \\ \hline
       4 & deg 1 &	deg 2 &	deg 1 &	0.8838 \\ \hline
       5 & deg 1 &	deg 2 &	deg 2 &	0.8978 \\ \hline
       6 & deg 1 &	deg 2 &	deg 3 &	0.8971 \\ \hline
       7 & deg 1 &	deg 3 &	deg 1 &	0.8937 \\ \hline
       8 & deg 1 &	deg 3 &	deg 2 &	0.9003 \\ \hline
       9 & deg 1 &	deg 3 &	deg 3 &	0.8994 \\ \hline
       10 & deg 2 &	deg 1 &	deg 1 &	0.9144 \\ \hline
       11 & deg 2 &	deg 1 &	deg 2 &	0.9292 \\ \hline
       12 & deg 2 &	deg 1 &	deg 3 &	0.9283 \\ \hline
       13 & deg 2 &	deg 2 &	deg 1 &	0.9154 \\ \hline
       14 & deg 2 &	deg 2 &	deg 2 &	0.9303 \\ \hline
       15 & deg 2 &	deg 2 &	deg 3 &	0.9294 \\ \hline
       16 & deg 2 &	deg 3 &	deg 1 &	0.9244 \\ \hline
       17 & deg 2 &	deg 3 &	deg 2 &	0.9323 \\ \hline
       18 & deg 2 &	deg 3 &	deg 3 &	0.9314 \\ \hline
       19 & deg 3 &	deg 1 &	deg 1 &	0.9136 \\ \hline
       20 & deg 3 &	deg 1 &	deg 2 &	0.9298 \\ \hline
       21 & deg 3 &	deg 1 &	deg 3 &	0.9287 \\ \hline
       22 & deg 3 &	deg 2 &	deg 1 &	0.9146 \\ \hline
       23 & deg 3 &	deg 2 &	deg 2 &	0.9307 \\ \hline
       24 & deg 3 &	deg 2 &	deg 3 &	0.9297 \\ \hline
       25 & deg 3 &	deg 3 &	deg 1 &	0.9235 \\ \hline
       26 & deg 3 &	deg 3 &	deg 2 &	0.9324 \\ \hline
       27 & deg 3 &	deg 3 &	deg 3 &	0.9314 \\ \hline
      \end{tabular}}
  \end{center}
\end{table}

As the increase in adjusted \textit{R}-squared is not significant, we adopt a linear model in which a linear relationship is assumed between the four variables. Travel Time on an edge is calculated using the formula given below.

Travel time on an edge = $\alpha$*(length of the edge) +  $\beta$*(number of vehicles on the edge) +  $\gamma$*(speed of slowest vehicle on the edge) +  $\delta$

Linear regression is applied to obtain the values of  $\alpha$, $\beta$, $\gamma$, and $\delta$. The software Stata 15.0 is used to run the regression. The obtained values of $\alpha$, $\beta$, $\gamma$, and $\delta$ are shown in Table \ref{tab:table3.777}.

\begin{table}[H]
  \begin{center}
   \caption{\\ Adjusted R-squared values of regression model with different degrees w.r.t. the independent variables}
    \label{tab:table3.777}
    \scalebox{0.9}{
    \begin{tabular}{rrrrrr}
    \hline
      Case & \textbf{$\alpha$} & \textbf{$\beta$} & \textbf{$\gamma$} & \textbf{$\delta$} & \textbf{Adjusted R-squared} \\\hline
       1 & 0.0410676 & 1.06991 &	-0.0521752 &	0.2179376 &	0.9642 \\ \hline
       2 & 0.0372134 &	0.4077467 &	-0.7931202 &	10.94339 &	0.9237 \\ \hline
       3 & 0.1133235 & 0.10954 &	-0.3618684 &	1.489052 &	0.9073 \\ \hline
       4 & 0.1091237 &	0.5725384 &	-4.477951 &	33.56182 &	0.8906 \\ \hline
       5 & 0.0554988 & 0.405355 &	-1.226947 &	18.81044 &	0.8854\\ \hline

      \end{tabular}}
  \end{center}
\end{table}

\begin{figure}[H]
\centering
\includegraphics[scale=0.5]{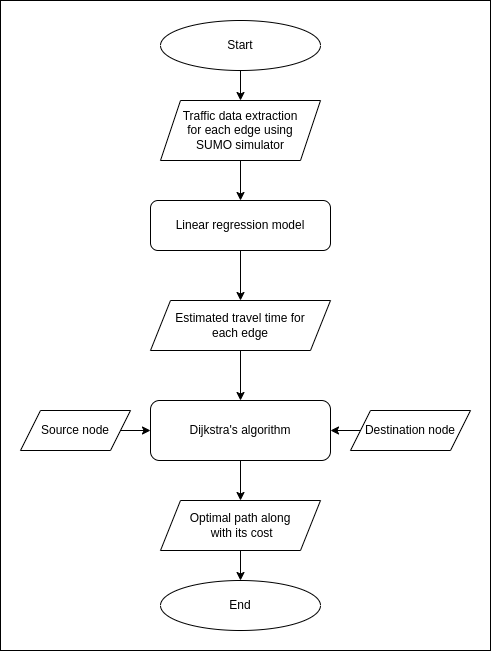}
\caption{Flowchart for determining the optimal path}
\label{image-t1}
\end{figure}

The developed regression model estimates the travel time on any edge with the given traffic conditions on the IIT Kharagpur network (see Table \ref{tab:table3.47}). Fig. \ref{image-t1} depicts the derivation of the optimal path between any two nodes.

\begin{table}[H]
  \begin{center}
   \caption{\\ Traffic under different conditions}
    \label{tab:table3.47}
    \scalebox{0.81}{
    \begin{tabular}{rrrr}
    \hline
      Case & Edge with number of lanes & Length of the edge (m) & \parbox[t]{4.5cm}{Speed of the slowest vehicle \\ on the edge (m/s)} \\\hline
       1 & $>$ 1 & $<$ 400 & not restricted \\ \hline
       2 & $>$ 1 &	$>$ 400 &	not restricted \\ \hline
       3 & 1 &	$<$ 100 &	not restricted \\ \hline
       4 & 1 &	$>$ 100 &	$<$ 10 \\ \hline
       5 & 1 & $>$ 100	& $>$ 10 \\ \hline
      \end{tabular}}
  \end{center}
\end{table}

Here, first, the traffic data is extracted for each edge. Then this data is fed to the regression model which estimates the travel time for each edge. These estimated travel times are used in Dijkstra's algorithm to obtain the optimal path and the corresponding travel cost.

At the start of a simulation, the crime node and the initial locations of the defender vehicles are chosen at random. Considering the SDVA problem, for $m$ defender vehicles, $m$ critical junctions are identified within 500 meters of the crime scene. For each defender, we calculate the travel cost of the optimal path in terms of the estimated travel time from the initial location to each critical junction. Hence, we obtain $m^2$ paths and their corresponding travel costs. We use the Hungarian algorithm to assign the $m$ critical junctions to the $m$ defender vehicles.

In the absence of traffic, we use Dijkstra's algorithm to determine the optimal path between any two nodes. Here the total travel cost of the optimal path depends on the shortest distance only.

Considering simulations in the presence of traffic, we add a total of 501 random (civilian) vehicles to the IIT Kharagpur network with speed factors uniformly distributed between 0.25 and 0.75. In the simulations, the defender vehicles are assigned to the critical junctions optimally. Then defenders travel to their respective junctions and halt there until a terminating condition is fulfilled. The terminating condition for success is when the distance between the attacker and any one of the defenders becomes less than 30 meters and the utility value of defenders becomes 1. In contrast, the terminating condition for failure is when the distance between the attacker and any one of the exit nodes becomes less than 30 meters and defenders get a utility of 0.

The process of generating the optimal path between any two nodes is shown in Fig. \ref{image-t1}. In case of DVRSP with traffic, a total of 501 random (civilian) vehicles are added to the IIT Kharagpur network with speed factors uniformly distributed between 0.25 and 0.75. A defender $d^r$ waits at a particular node $v^r$ in his path till the $t^{r,out}$ value associated with that node. Again, like the SDVA, the game runs until one of the earlier mentioned termination conditions is fulfilled.

We generate 20 random instances with a fixed number of defender vehicles. An instance basically refers to a pair consisting of the crime node and the initial locations of the defenders. We report the average utility at the end of the 20 simulations. This average utility is used as the performance measure. Here, an average utility of 0.4 indicates that 40\% of the time, the defenders are successful in interdicting the attacker. This can also be called the success rate of the defenders. The process of running the 20 simulations is repeated for both approaches while the number of defender vehicles varies from one to four.

\section{Results and discussion}
\label{S:6}
In this section, we present the result of the proposed approaches. The proposed algorithms are coded in python 2.7 and tested on an Intel(R) Core(TM) 3.20 GHz processor with 4 GB RAM PC under LINUX environment. We collect the dataset using the SUMO simulator. On the IIT Kharagpur map, India, the urban road network can be defined as a graph with 100 junctions (nodes) and 246 roads (edges). SUMO provides the edge lengths. In the graph, the crime scene can be any node. There are five fixed exit nodes. Random nodes, except the exit nodes and crime scene node, are the initial location of defenders. In the algorithm named SDVASE, the value of $d_{min}$ is set to 15 meters and $distance_{min}$ is set to 2 KM. The value of $max_{iterations}$ is set to 30. Considering GA, the value of $K_{max}$ is also set to 30. We show that GA performs better than MILP for defender (bestDO) developed by Zhang et al. \cite{zhang2017optimal}. BestDo is an optimization model in which the optimal formulation is done for defender movements over time. If more attacker paths are intercepted by new defender paths, then defenders get a higher utility. As MILP does not converge for large netwroks, we show the performance evaluation of the proposed GA with smaller size grids. Considering grids of size 5*5 to 9*9, GA provides the best solution like MILP (bestDO) but takes significantly less time than MILP. We have provided the details in Table \ref{tab:tables0}.

\begin{table}[H]
  \begin{center}
   \caption{\\ Computational results of GA vs. MILP (bestDO)}
    \label{tab:tables0}
    \scalebox{0.75}{
    \begin{tabular}{rrrrrrrr}
    \hline
      \parbox[t]{1.0cm}{Size of \\ network} & \parbox[t]{0.8cm}{Crime \\ node} & \parbox[t]{1.0cm}{Exit \\ points} & \parbox[t]{2.5cm}{Initial defender \\ locations} & \parbox[t]{1.0cm}{Utility \\ of GA} & \parbox[t]{1.2cm}{Utility \\ of MILP} & \parbox[t]{2.0cm}{Run time \\ of GA (sec)} & \parbox[t]{2.0cm}{Run time \\ of MILP (sec)}\\ \hline
       5*5 & 13 & 1, 16, 20 & 3, 15 & 1.0 & 1.0 & 1.37 & 14.99\\ \hline
       6*6 & 22 & 3, 6, 35 & 4, 34 & 1.0 & 1.0 & 1.93 & 63.93\\ \hline
       7*7 & 25 & 3, 11, 35 & 14, 45 & 1.0 & 1.0 & 2.44 & 72.31\\ \hline
       8*8 & 37 & 17, 48, 57 & 5, 33 & 1.0 & 1.0 & 2.69 & 87.83\\ \hline
       9*9 & 41 & 4, 28, 37, 40, 72 & 35, 77 & 1.0 & 1.0 & 5.76 & 109.33\\ \hline
       \end{tabular}}
  \end{center}
\end{table}

In Table \ref{tab:table453}, we consider the without traffic scenario, and we show that the developed DVRSP model performs better than the SDVA. Considering the computational time, DVRSP can solve the problem in a few seconds, while SDVA takes more time compared to DVRSP. In addition, solution quality is better considering DVRSP. Here, we report 25 test cases, showing that the interdiction probability is better in case of DVRSP.

\begin{table}[H]
  \begin{center}
   \caption{Computational results of the developed SDVA model and DVRSP model on the IIT Kharagpur map}
    \label{tab:table453}
    \scalebox{0.51}{
    \begin{tabular}{lllllllllll} 
    \hline
      \parbox[t]{1.2cm}{\textbf{Test \\ Case}} & \multicolumn{10}{c}{\textbf{Network size: 100 nodes, 246 edges}} \\
      \cline{2-11}
      &  \parbox[t]{1.5cm}{Crime \\ Location} & \parbox[t]{1.5cm}{Initial \\ police\\  Locations} & \parbox[t]{1.5cm}{Exit \\ Point} & \parbox[t]{1.5cm}{Halting \\ Locations} & \parbox[t]{1.7cm}{SDVA \\ Time (Sec)} & \parbox[t]{1.7cm}{SDVA \\ Utility} & \parbox[t]{1.7cm}{SDVA \\ Result} & \parbox[t]{1.3cm}{DVRSP \\ Time (Sec)} & \parbox[t]{1.3cm}{DVRSP \\ Utility} & \parbox[t]{1.3cm}{DVRSP \\ Result} \\ \hline
       1 & 5 & 19 & 44 & 15 & 33 & 1 & 40\% & 1.56 & 0.34 & 44.20\%\\ \cline{1-7}\cline{9-10}
       2 & 39 & 33 & 100 & 63 & 112 & 0 & & 1.62 & 0.17 & \\ \cline{1-7}\cline{9-10}
       3 & 65 & 50 & 100 & 36 & 57 & 1 & & 1.53 & 0.68 & \\ \cline{1-7}\cline{9-10}
       4 & 28 & 27 & 64 & 5 & 52 & 0 & & 1.55 & 0.51 & \\ \cline{1-7}\cline{9-10}
       5 & 52 & 48 & 96 & 82 & 39 & 0 & & 1.59 & 0.51 & \\ \hline
       6 & 27 & [11, 63] & 44 & [5, 29] & 55 & 1 & 60\% & 3.11 & 0.34 & 61.20\%\\ \cline{1-7}\cline{9-10}
       7 & 26 & [22, 52] & 96 & [24, 27] & 59 & 0 & & 3.12 & 0.68 & \\ \cline{1-7}\cline{9-10}
       8 & 99 & [59, 53] & 2 & [75, 94] & 69 & 1 & & 3.15 & 0.68 & \\ \cline{1-7}\cline{9-10}
       9 & 10 & [54, 83] & 44 & [7, 36] & 40 & 1 & & 2.97 & 0.85 & \\ \cline{1-7}\cline{9-10}
       10 & 38 & [41, 36] & 9 & [63, 39] & 106 & 0 & & 2.99 & 0.51 & \\ \hline
       11 & 30 & [46, 6, 33] & 44 & [15, 31, 5] & 38 & 1 & 40\% & 5.52 & 0.85 & 77.40\%\\ \cline{1-7}\cline{9-10}
       12 & 61 & [68, 19, 32] & 64 & [15, 16, 31] & 30 & 1 & & 4.54 & 0.68 & \\ \cline{1-7}\cline{9-10}
       13 & 87 & [74, 18, 49] & 9 & [46, 48, 44] & 46 & 0 & & 2.36 & 0.99 & \\ \cline{1-7}\cline{9-10}
       14 & 24 & [97, 61, 46] & 64 & [15, 28, 5] & 163 & 0 & & 4.52 & 0.85 & \\ \cline{1-7}\cline{9-10}
       15 & 75 & [78, 43, 36] & 100 & [48, 44, 46] & 101 & 0 & & 1.65 & 0.5 & \\ \hline
       16 & 68 & [58, 59, 94, 21] & 9 & [62, 66, 83, 48] & 70 & 1 & 60\% & 2.17 & 1 & 69.80\%\\ \cline{1-7}\cline{9-10}
       17 & 43 & [3, 69, 1, 99] & 9 & [52, 47, 45, 49] & 123 & 0 & & 3.1 & 0.33 & \\ \cline{1-7}\cline{9-10}
       18 & 94 & [81, 22, 56, 73] & 96 & [99, 75, 2, 6] & 27 & 0 & & 2.22 & 1 & \\ \cline{1-7}\cline{9-10}
       19 & 10 & [83, 72, 42, 76] & 96 & [6, 7, 36, 15] & 66 & 1 & & 2.04 & 0.5 & \\ \cline{1-7}\cline{9-10}
       20 & 31 & [32, 38, 57, 98] & 64 & [15, 6, 53, 2] & 40 & 1 & & 3.19 & 0.66 & \\ \hline
       21 & 42 & [30, 89, 19, 58, 82] & 64 & [23, 61, 62, 16, 48] & 47 & 1 & 80\% & 2.84 & 1 & 90\%\\ \cline{1-7}\cline{9-10}
       22 & 12 & [70, 56, 71, 65, 98] & 64 & [6, 7, 36, 10, 15] & 53 & 1 & & 2.63 & 1 & \\ \cline{1-7}\cline{9-10}
       23 & 98 & [51, 79, 17, 88, 82] & 100 & [100, 94, 75, 95, 93] & 47 & 0 & & 2.74 & 0.5 & \\ \cline{1-7}\cline{9-10}
       24 & 32 & [26, 74, 25, 28, 69] & 9 & [41, 5, 33, 6, 42] & 51 & 1 & & 2.56 & 1 & \\ \cline{1-7}\cline{9-10}
       25 & 41 & [91, 84, 31, 22, 3] & 2 & [5, 6, 15, 42, 16] & 63 & 1 & & 2.64 & 1 & \\ \hline
    \end{tabular}}
  \end{center}
\end{table}

\begin{figure}[H]
\centering
\includegraphics[scale=0.5]{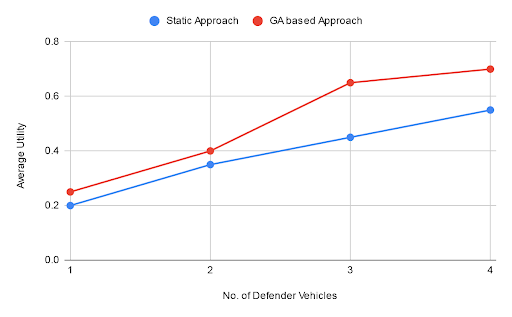}
\caption{Comparison of SDVA and DVRSP in the absence of traffic}
\label{image-j11}
\end{figure}

To compare the performance of the SDVA and DVRSP problems, we run simulations using the SUMO simulator both in the absence and presence of traffic in the network. In case of without traffic in the network, the developed approach for SDVA shows an average utility of 0.2 for only 1 defender, whereas the average utility is 0.55 for 4 defenders. In case of DVRSP, the developed GA-based meta-heuristic approach shows better results than the previous approach. Here we get an average utility of 0.25 considering only 1 defender and 0.7 with 4 defenders. The results of the simulated instances without traffic are presented in Tables \ref{tab:tables1}, \ref{tab:tables2}, \ref{tab:tables3}, and \ref{tab:tables4}, with the number of defenders varying from 1 to 4, respectively. The final results of the two approaches without traffic are shown in Table \ref{tab:tables5}. The comparison of the two approaches is shown in Fig. \ref{image-j11}.

\begin{table}[H]
  \begin{center}
   \caption{\\ Results of simulated instances for 1 defender vehicle in the absence of traffic}
    \label{tab:tables1}
    \scalebox{0.45}{
    \begin{tabular}{rrrrr}
    \hline
      Test case & Crime node & Police station & \textbf{Utility of SDVA} & \textbf{Utility of DVRSP} \\\hline
       1 & 5 & 26 &	0 &	1 \\ \hline
       2 & 63 &	19 & 0 & 0 \\ \hline
       3 & 49 &	20 & 1 & 1 \\ \hline
       4 & 17 &	56 & 0 & 0 \\ \hline
       5 & 72 & 34 & 0 & 0 \\ \hline
       6 & 65 & 42 & 0 & 0 \\ \hline
       7 & 48 & 31 & 0 & 0 \\ \hline
       8 & 3  & 14 & 0 & 1 \\ \hline
       9 & 29 & 49 & 0 & 0 \\ \hline
       10 & 67 & 62 & 1 & 1 \\ \hline
       11 & 92 & 35 & 0 & 0 \\ \hline
       12 & 53 & 82 & 1 & 0 \\ \hline
       13 & 28 & 42 & 0 & 0 \\ \hline
       14 & 41 & 63 & 0 & 0 \\ \hline
       15 & 20 & 59 & 0 & 0 \\ \hline
       16 & 3 & 21 & 1 & 1 \\ \hline
       17 & 75 & 67 & 0 & 0 \\ \hline
       18 & 91 & 49 & 0 & 0 \\ \hline
       19 & 50 & 67 & 0 & 0 \\ \hline
       20 & 33 & 52 & 0 & 0 \\ \hline
       \end{tabular}}
  \end{center}
\end{table}

\begin{table}[H]
  \begin{center}
   \caption{\\ Results of simulated instances for 2 defender vehicles in the absence of traffic}
    \label{tab:tables2}
    \scalebox{0.45}{
    \begin{tabular}{rrrrr}
    \hline
      Test case & Crime node & Police station & \textbf{Utility of SDVA} & \textbf{Utility of DVRSP} \\\hline
       1 & 2 & 33, 56 & 1 & 0 \\ \hline
       2 & 43 & 23, 16 & 1 & 1 \\ \hline
       3 & 59 & 42, 86 & 0 & 0 \\ \hline
       4 & 52 & 23, 60 & 0 & 0 \\ \hline
       5 & 32 & 11, 46 & 0 & 0 \\ \hline
       6 & 65 & 42, 79 & 0 & 1 \\ \hline
       7 & 79 & 51, 95 & 0 & 0 \\ \hline
       8 & 83 & 25, 34 & 0 & 0 \\ \hline
       9 & 94 & 51, 49 & 1 & 1 \\ \hline
       10 & 16 & 37, 55 & 1 & 1 \\ \hline
       11 & 24 & 30, 66 & 1 & 1 \\ \hline
       12 & 12 & 56, 37 & 1 & 1 \\ \hline
       13 & 33 & 39, 51 & 0 & 0 \\ \hline
       14 & 5 & 52, 24 & 1 & 1 \\ \hline
       15 & 46 & 49, 70 & 0 & 0 \\ \hline
       16 & 83 & 40, 59 & 0 & 0 \\ \hline
       17 & 51 & 41, 95 & 0 & 0 \\ \hline
       18 & 43 & 22, 50 & 0 & 0 \\ \hline
       19 & 55 & 10, 26 & 0 & 1 \\ \hline
       20 & 61 & 8, 97 & 0 & 0 \\ \hline

       \end{tabular}}
  \end{center}
\end{table}

\begin{table}[H]
  \begin{center}
   \caption{\\ Results of simulated instances for 3 defender vehicles in the absence of traffic}
    \label{tab:tables3}
    \scalebox{0.45}{
    \begin{tabular}{rrrrr}
    \hline
      Test case & Crime node & Police station & \textbf{Utility of SDVA} & \textbf{Utility of DVRSP} \\\hline
       1 & 69 & 45, 68, 82 & 1 & 1 \\ \hline
       2 & 51 & 15, 65, 43 & 1 & 1 \\ \hline
       3 & 32 & 53, 48, 97 & 0 & 0 \\ \hline
       4 & 54 & 30, 42, 60 & 1 & 1 \\ \hline
       5 & 12 & 25, 86, 53 & 1 & 1 \\ \hline
       6 & 8 & 89, 75, 22 & 1 & 1 \\ \hline
       7 & 75 & 15, 56, 48 & 0 & 0 \\ \hline
       8 & 43 & 23, 46, 85 & 0 & 1 \\ \hline
       9 & 22 & 56, 99, 63 & 1 & 1 \\ \hline
       10 & 4 & 34, 59, 20 & 0 & 0 \\ \hline
       11 & 13 & 30, 45, 51 & 0 & 1 \\ \hline
       12 & 36 & 4, 49, 57 & 0 & 1 \\ \hline
       13 & 71 & 12, 58, 33 & 0 & 0 \\ \hline
       14 & 53 & 22, 40, 81 & 0 & 1 \\ \hline
       15 & 95 & 52, 40, 5 & 0 & 0 \\ \hline
       16 & 50 & 61, 40, 73 & 0 & 1 \\ \hline
       17 & 40 & 52, 94, 30 & 1 & 1 \\ \hline
       18 & 29 & 60, 37, 49 & 1 & 0 \\ \hline
       19 & 10 & 35, 68, 42 & 1 & 0 \\ \hline
       20 & 8 & 38, 47, 77 & 0 & 1 \\ \hline
       \end{tabular}}
  \end{center}
\end{table}

\begin{table}[H]
  \begin{center}
   \caption{\\ Results of simulated instances for 4 defender vehicles in the absence of traffic}
    \label{tab:tables4}
    \scalebox{0.45}{
    \begin{tabular}{rrrrr}
    \hline
      Test case & Crime node & Police station & \textbf{Utility of SDVA} & \textbf{Utility of DVRSP} \\\hline
       1 & 19 & 12, 53, 47, 79 & 0 & 0 \\ \hline
       2 & 56 & 2, 59,  40, 61 & 1 & 1 \\ \hline
       3 & 42 & 99, 68, 15, 34 & 1 & 0 \\ \hline
       4 & 57 & 27, 49, 26, 14 & 0 & 1 \\ \hline
       5 & 20 & 35, 23, 59, 24 & 0 & 1 \\ \hline
       6 & 34 & 67, 19, 31, 3 & 1 & 1 \\ \hline
       7 & 41 & 81, 32, 99, 56 & 1 & 0 \\ \hline
       8 & 63 & 72, 28, 53, 27 & 1 & 0 \\ \hline
       9 & 60 & 31, 33, 48, 61 & 1 & 0 \\ \hline
       10 & 90 & 2, 63, 80, 92 & 0 & 1 \\ \hline
       11 & 71 & 12, 20, 75, 22 & 1 & 1 \\ \hline
       12 & 84 & 68, 3, 82, 72 & 0 & 1 \\ \hline
       13 & 77 & 43, 50, 1, 29 & 0 & 0 \\ \hline
       14 & 49 & 42, 17, 22, 16 & 1 & 1 \\ \hline
       15 & 80 & 70, 72, 88, 46 & 0 & 1 \\ \hline
       16 & 3 & 83, 19, 43, 24 & 1 & 1 \\ \hline
       17 & 39 & 37, 31, 85, 27 & 0 & 1 \\ \hline
       18 & 47 & 46, 39, 53, 49 & 1 & 1 \\ \hline
       19 & 52 & 91, 55, 84, 19 & 0 & 1 \\ \hline
       20 & 11 & 63, 29, 69, 74 & 1 & 1 \\ \hline
       \end{tabular}}
  \end{center}
\end{table}

\begin{table}[H]
  \begin{center}
   \caption{\\ Comparison of average utilities of both approaches in the absence of traffic}
    \label{tab:tables5}
    \scalebox{0.5}{
    \begin{tabular}{rrrr}
    \hline
      Test case & Number of defender vehicles & \textbf{Average utility of SDVA} & \textbf{ Average utility of DVRSP} \\\hline
       1 & 1 & 0.2 & 0.25 \\ \hline
       2 & 2 & 0.35 & 0.4 \\ \hline
       3 & 3 & 0.45 & 0.65 \\ \hline
       4 & 4 & 0.55 & 0.7 \\ \hline
       \end{tabular}}
  \end{center}
\end{table}

In case of SDVA with traffic, we report 25 test cases and show that the interdiction probability increases with the increase of traffic. The result is shown in the Table \ref{tab:table53}. We calculate the probability of interdiction by varying the initial locations of police, number of police vehicles, crime scene, and exit point.

\begin{table}[H]
  \begin{center}
   \caption{Computational results of the developed SDVA model without traffic and with traffic }
    \label{tab:table53}
    \scalebox{0.51}{
    \begin{tabular}{lllllllllll} 
    \hline
      \textbf{Test case} & \multicolumn{10}{c}{\textbf{Network size: 100 nodes, 246 edges}} \\
      \cline{2-11}
      &  \parbox[t]{1.5cm}{Crime \\ location} & \parbox[t]{1.5cm}{Initial \\ police\\  locations} & \parbox[t]{1.5cm}{Exit \\ point} & \parbox[t]{1.5cm}{Halting \\ locations} & \parbox[t]{1.5cm}{Time (Sec) \\(without \\ traffic)} & \parbox[t]{1.5cm}{Utility \\(without \\ traffic)} & \parbox[t]{1.5cm}{Result \\(without \\ traffic)} & \parbox[t]{1.5cm}{Time (Sec) \\(with \\ traffic)} & \parbox[t]{1.5cm}{Utility \\(with \\ traffic)} & \parbox[t]{1.5cm}{Result \\(with \\ traffic)} \\ \hline
       1 & 5 & 19 & 44 & 15 & 33 & 1 & 40\% & 439 & 1 & 80\%\\ \cline{1-7}\cline{9-10}
       2 & 39 & 33 & 100 & 63 & 112 & 0 & & 304 & 1 & \\ \cline{1-7}\cline{9-10}
       3 & 65 & 50 & 100 & 36 & 57 & 1 & & 464 & 0 & \\ \cline{1-7}\cline{9-10}
       4 & 28 & 27 & 64 & 5 & 52 & 0 & & 499 & 1 & \\ \cline{1-7}\cline{9-10}
       5 & 52 & 48 & 96 & 82 & 39 & 0 & & 499 & 1 & \\ \hline
       6 & 27 & [11, 63] & 44 & [5, 29] & 55 & 1 & 60\% & 320 & 1 & 60\%\\ \cline{1-7}\cline{9-10}
       7 & 26 & [22, 52] & 96 & [24, 27] & 59 & 0 & & 230 & 0 & \\ \cline{1-7}\cline{9-10}
       8 & 99 & [59, 53] & 2 & [75, 94] & 69 & 1 & & 123 & 1 & \\ \cline{1-7}\cline{9-10}
       9 & 10 & [54, 83] & 44 & [7, 36] & 40 & 1 & & 259 & 1 & \\ \cline{1-7}\cline{9-10}
       10 & 38 & [41, 36] & 9 & [63, 39] & 106 & 0 & & 299 & 0 & \\ \hline
       11 & 30 & [46, 6, 33] & 44 & [15, 31, 5] & 38 & 1 & 40\% & 499 & 1 & 60\%\\ \cline{1-7}\cline{9-10}
       12 & 61 & [68, 19, 32] & 64 & [15, 16, 31] & 30 & 1 & & 202 & 1 & \\ \cline{1-7}\cline{9-10}
       13 & 87 & [74, 18, 49] & 9 & [46, 48, 44] & 46 & 0 & & 343 & 0 & \\ \cline{1-7}\cline{9-10}
       14 & 24 & [97, 61, 46] & 64 & [15, 28, 5] & 163 & 0 & & 499 & 1 & \\ \cline{1-7}\cline{9-10}
       15 & 75 & [78, 43, 36] & 100 & [48, 44, 46] & 101 & 0 & & 141 & 0 & \\ \hline
       16 & 68 & [58, 59, 94, 21] & 9 & [62, 66, 83, 48] & 70 & 1 & 60\% & 499 & 1 & 80\%\\ \cline{1-7}\cline{9-10}
       17 & 43 & [3, 69, 1, 99] & 9 & [52, 47, 45, 49] & 123 & 0 & & 499 & 1 & \\ \cline{1-7}\cline{9-10}
       18 & 94 & [81, 22, 56, 73] & 96 & [99, 75, 2, 6] & 27 & 0 & & 83 & 0 & \\ \cline{1-7}\cline{9-10}
       19 & 10 & [83, 72, 42, 76] & 96 & [6, 7, 36, 15] & 66 & 1 & & 312 & 1 & \\ \cline{1-7}\cline{9-10}
       20 & 31 & [32, 38, 57, 98] & 64 & [15, 6, 53, 2] & 40 & 1 & & 207 & 1 & \\ \hline
       21 & 42 & [30, 89, 19, 58, 82] & 64 & [23, 61, 62, 16, 48] & 47 & 1 & 80\% & 217 & 1 & 80\%\\ \cline{1-7}\cline{9-10}
       22 & 12 & [70, 56, 71, 65, 98] & 64 & [6, 7, 36, 10, 15] & 53 & 1 & & 190 & 1 & \\ \cline{1-7}\cline{9-10}
       23 & 98 & [51, 79, 17, 88, 82] & 100 & [100, 94, 75, 95, 93] & 47 & 0 & & 138 & 0 & \\ \cline{1-7}\cline{9-10}
       24 & 32 & [26, 74, 25, 28, 69] & 9 & [41, 5, 33, 6, 42] & 51 & 1 & & 326 & 1 & \\ \cline{1-7}\cline{9-10}
       25 & 41 & [91, 84, 31, 22, 3] & 2 & [5, 6, 15, 42, 16] & 63 & 1 & & 329 & 1 & \\ \hline
    \end{tabular}}
  \end{center}
\end{table}

\begin{figure}[H]
\centering
\includegraphics[scale=0.5]{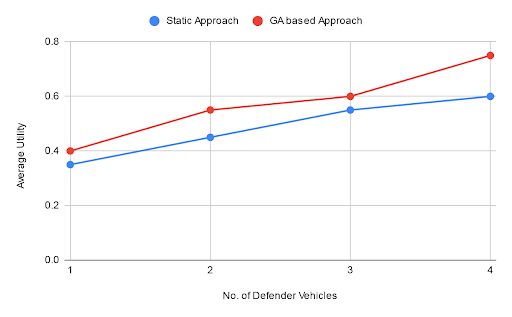}
\caption{Comparison of SDVA and DVRSP in the presence of traffic}
\label{image-j12}
\end{figure}

As the interdiction probability increases with the increase of traffic in case of SDVA, we focus on comparing the performance of SDVA with DVRSP in presence of traffic. In case of the traffic on the IIT Kharagpur network, the approach for SDVA shows an average utility of 0.35 with only 1 defender and a utility of 0.6 with 4 defenders. The GA-based approach for DVRSP shows an average utility of 0.4 with only 1 defender and a utility of 0.75 with 4 defenders. The results of the simulated instances with traffic are presented in Tables \ref{tab:tables6}, \ref{tab:tables7}, \ref{tab:tables8}, and \ref{tab:tables9}, where the number of defender vehicles is equal to 1, 2, 3, and 4, respectively. The final results of the two approaches are tabulated in Table \ref{tab:tables10}. We show the comparison between the two strategies in Fig. \ref{image-j12}.

\begin{table}[H]
  \begin{center}
   \caption{\\ Results of simulated instances for 1 defender vehicle in the presence of traffic}
    \label{tab:tables6}
    \scalebox{0.5}{
    \begin{tabular}{rrrrr}
    \hline
      Test case & Crime node & Police station & \textbf{Utility of SDVA} & \textbf{Utility of DVRSP} \\\hline
       1 & 46 & 37 & 0 & 0 \\ \hline
       2 & 28 & 85 & 1 & 1 \\ \hline
       3 & 13 & 32 & 0 & 0 \\ \hline
       4 & 31 & 38 & 0 & 1 \\ \hline
       5 & 29 & 8 & 1 & 1 \\ \hline
       6 & 5 & 21 & 0 & 0 \\ \hline
       7 & 65 & 76 & 0 & 0 \\ \hline
       8 & 70 & 33 & 1 & 1 \\ \hline
       9 & 98 & 20 & 0 & 0 \\ \hline
       10 & 54 & 47 & 0 & 0 \\ \hline
       11 & 79 & 35 & 1 & 1 \\ \hline
       12 & 63 & 28 & 0 & 0 \\ \hline
       13 & 84 & 48 & 0 & 0 \\ \hline
       14 & 14 & 65 & 1 & 1 \\ \hline
       15 & 72 & 48 & 1 & 1 \\ \hline
       16 & 39 & 28 & 0 & 0 \\ \hline
       17 & 19 & 38 & 0 & 0 \\ \hline
       18 & 86 & 93 & 0 & 0 \\ \hline
       19 & 95 & 60 & 0 & 0 \\ \hline
       20 & 90 & 2 & 1 & 1 \\ \hline
       \end{tabular}}
  \end{center}
\end{table}

\begin{table}[H]
  \begin{center}
   \caption{\\ Results of simulated instances for 2 defender vehicles in the presence of traffic}
    \label{tab:tables7}
    \scalebox{0.5}{
    \begin{tabular}{rrrrr}
    \hline
      Test case & Crime node & Police station & \textbf{Utility of SDVA} & \textbf{Utility of DVRSP} \\\hline
       1 & 31 & 59, 19 & 1 & 0 \\ \hline
       2 & 99 & 27, 3 & 0 & 0 \\ \hline
       3 & 22 & 43, 56 & 1 & 1 \\ \hline
       4 & 23 & 6, 93 & 1 & 1 \\ \hline
       5 & 21 & 47, 82 & 0 & 0 \\ \hline
       6 & 25 & 76, 17 & 0 & 1 \\ \hline
       7 & 25 & 61, 46 & 0 & 1 \\ \hline
       8 & 80 & 71, 72 & 1 & 1 \\ \hline
       9 & 33 & 17, 28 & 1 & 1 \\ \hline
       10 & 90 & 30, 36 & 0 & 1 \\ \hline
       11 & 68 & 32, 36 & 0 & 1 \\ \hline
       12 & 41 & 82, 84 & 0 & 0 \\ \hline
       13 & 29 & 46, 75 & 1 & 0 \\ \hline
       14 & 24 & 55, 90 & 1 & 1 \\ \hline
       15 & 38 & 48, 77 & 0 & 1 \\ \hline
       16 & 16 & 78, 25 & 0 & 0 \\ \hline
       17 & 90 & 28, 65 & 0 & 0 \\ \hline
       18 & 94 & 41, 65 & 1 & 0 \\ \hline
       19 & 8 & 39, 6 & 1 & 1 \\ \hline
       20 & 39 & 10, 79 & 0 & 0 \\ \hline
       \end{tabular}}
  \end{center}
\end{table}

\begin{table}[H]
  \begin{center}
   \caption{\\ Results of simulated instances for 3 defender vehicles in the presence of traffic}
    \label{tab:tables8}
    \scalebox{0.5}{
    \begin{tabular}{rrrrr}
    \hline
      Test case & Crime node & Police station & \textbf{Utility of SDVA} & \textbf{Utility of DVRSP} \\\hline
       1 & 52 & 41, 51, 99 & 0 & 0 \\ \hline
       2 & 50 & 76, 47, 84 & 0 & 1 \\ \hline
       3 & 60 & 53, 20, 90 & 0 & 0 \\ \hline
       4 & 22 & 75, 15, 66 & 1 & 1 \\ \hline
       5 & 41 & 83, 27, 46 & 1 & 0 \\ \hline
       6 & 5 & 93, 19, 21 & 1 & 1 \\ \hline
       7 & 30 & 59, 82, 5 & 1 & 0 \\ \hline
       8 & 83 & 38, 18, 12 & 0 & 1 \\ \hline
       9 & 41 & 85, 76, 38 & 1 & 1 \\ \hline
       10 & 13 & 12, 73, 19 & 1 & 1 \\ \hline
       11 & 37 & 29, 68, 56 & 1 & 1 \\ \hline
       12 & 60 & 43, 98, 12 & 1 & 0 \\ \hline
       13 & 62 & 68, 47, 11 & 0 & 1 \\ \hline
       14 & 89 & 95, 99, 41 & 0 & 0 \\ \hline
       15 & 23 & 21, 88, 39 & 0 & 0 \\ \hline
       16 & 51 & 21, 47, 7 & 1 & 1 \\ \hline
       17 & 83 & 31, 46, 79 & 0 & 1 \\ \hline
       18 & 80 & 81, 22, 35 & 1 & 1 \\ \hline
       19 & 81 & 57, 63, 42 & 0 & 0 \\ \hline
       20 & 61 & 68, 65, 51 & 1 & 1 \\ \hline
       \end{tabular}}
  \end{center}
\end{table}

\begin{table}[H]
  \begin{center}
   \caption{\\ Results of simulated instances for 4 defender vehicles in the presence of traffic}
    \label{tab:tables9}
    \scalebox{0.5}{
    \begin{tabular}{rrrrr}
    \hline
      Test case & Crime node & Police station & \textbf{Utility of SDVA} & \textbf{Utility of DVRSP} \\\hline
       1 & 85 & 75, 65, 51, 53 & 1 & 1 \\ \hline
       2 & 22 & 19, 51, 12, 61 & 1 & 1 \\ \hline
       3 & 12 & 72, 77, 73, 40 & 1 & 0 \\ \hline
       4 & 80 & 72, 77, 73, 40 & 1 & 1 \\ \hline
       5 & 32 & 11, 85, 63, 61 & 1 & 1 \\ \hline
       6 & 55 & 77, 89, 86, 80 & 0 & 0 \\ \hline
       7 & 15 & 1, 85, 68, 77 & 0 & 1 \\ \hline
       8 & 81 & 74, 7, 26, 50 & 0 & 0 \\ \hline
       9 & 54 & 28, 27, 5, 66 & 0 & 1 \\ \hline
       10 & 73 & 79, 28, 59, 67 & 1 & 1 \\ \hline
       11 & 58 & 3, 15, 86, 5 & 1 & 1 \\ \hline
       12 & 39 & 99, 73, 78, 25 & 0 & 0 \\ \hline
       13 & 36 & 26, 85, 50, 45 & 0 & 1 \\ \hline
       14 & 76 & 49, 7, 14, 15 & 1 & 1 \\ \hline
       15 & 41 & 94, 37, 48, 23 & 1 & 1 \\ \hline
       16 & 46 & 32, 4, 25, 77 & 0 & 1 \\ \hline
       17 & 78 & 84, 57, 4, 32 & 1 & 1 \\ \hline
       18 & 42 & 69, 65, 90, 89 & 1 & 1 \\ \hline
       19 & 70 & 60, 75, 5, 18 & 1 & 1 \\ \hline
       20 & 58 & 45, 47, 35, 91 & 0 & 0 \\ \hline
       \end{tabular}}
  \end{center}
\end{table}

\begin{table}[H]
  \begin{center}
   \caption{\\ Comparison of average utilities of both approaches in the presence of traffic}
    \label{tab:tables10}
    \scalebox{0.5}{
    \begin{tabular}{rrrr}
    \hline
      Test case & Number of defender vehicles & \textbf{Average utility of SDVA} & \textbf{ Average utility of DVRSP} \\\hline
       1 & 1 & 0.35 & 0.4 \\ \hline
       2 & 2 & 0.45 & 0.55 \\ \hline
       3 & 3 & 0.55 & 0.6 \\ \hline
       4 & 4 & 0.6 & 0.75 \\ \hline
       \end{tabular}}
  \end{center}
\end{table}

Here, we see that the average utility increases with the increase in the number of resources, i.e., the number of defender vehicles. Also, it is evident that the developed GA-based approach for DVRSP performs better than the developed approach for SDVA both in the presence of traffic as well as in the absence of traffic in the network.  

\section{Conclusion}
\label{S:7}

In this paper, we develop an optimal resource allocation approach and a meta-heuristic to solve the escape interdiction problem in a simulation-based environment. In addition, we develop a novel linear regression model to find out the optimal path in presence of traffic.

Considering SDVA, we develop optimal resource allocation model in a large transportation network. Though this optimal resource allocation model for SDVA is scalable, we focus on developing a more advanced search strategy like routing with optimal resource allocation.
The GA-based meta-heuristic approach provides defender strategies for large transportation networks for the DVRSP problem. With a minimum number of defenders, the developed algorithm can solve large-scale problems within a short amount of time. Here we get a marginal improvement in the solution quality compared to the developed approach for SDVA. The proposed approach for DVRSP provides better interdiction probability because of the more city coverage.

There are a few limitations of the developed models. For example, the proposed methods do not consider related information that the defending unit can acquire during the interdiction operations except the initial crime node location. Hence, future research can be done to develop more sophisticated heuristic approach in simulation platform.

{}

\begin{thebibliography}{}

\bibitem{samanta2021literature}Samanta, Sukanya, Sen Goutam, and Ghosh Soumya Kanti. ``A literature review on police patrolling problems." Annals of Operations Research 316, no. 2 (2022): 1063-1106.

\bibitem{samanta2022vns}Samanta, Sukanya, Mohandass Tushar, Sen Goutam, and Ghosh Soumya Kanti. ``A VNS-based metaheuristic approach for escape interdiction on transportation networks." Computers \& Industrial Engineering 169, (2022): 108253.

\bibitem{samanta2021vehicle}Samanta, Sukanya, Sen Goutam, and Ghosh Soumya Kanti. ``Vehicle Interdiction Strategy in Complex Road Networks-A Simulation Based Approach." 2021 IEEE International Conference on Industrial Engineering and Engineering Management (IEEM) (2021): 1299-1302.

\bibitem{zhang2017optimal}Zhang, Youzhi, An Bo, Tran-Thanh Long, Wang Zhen, Gan Jiarui, and Jennings, Nicholas R. ``Optimal escape interdiction on transportation networks." Proceedings of the Twenty-Sixth International Joint Conference on Artificial Intelligence (2017): 3936-3944.

\bibitem{whitley1994genetic}Whitley, Darrell. ``A genetic algorithm tutorial." Statistics and computing 4, no. 2 (1994): 65-85.

\bibitem{harik1999compact}Harik, Georges R, Lobo Fernando G, and Goldberg David E. ``The compact genetic algorithm." IEEE transactions on evolutionary computation 3, no. 4 (1999): 287-297.

\bibitem{thangiah1993vehicle}Thangiah, Sam R. ``Vehicle routing with time windows using genetic algorithms." Citeseer (1993).

\bibitem{paruchuri2007efficient}Paruchuri, Praveen, Pearce Jonathan P, Tambe Milind, Ordonez Fernando, and Kraus Sarit. ``An efficient heuristic approach for security against multiple adversaries." Proceedings of the 6th international joint conference on Autonomous agents and multiagent systems (2007): 1-8, \url{https://doi.org/10.1145/1329125.1329344}.

\bibitem{bier2008optimal}Bier, Vicki M, Haphuriwat Naraphorn, Menoyo Jaime, Zimmerman Rae, and Culpen Alison M. ``Optimal resource allocation for defense of targets based on differing measures of attractiveness." Risk Analysis: An International Journal 28, no. 3 (2008): 763-770.

\bibitem{tsai2009iris}Tsai, Jason, Rathi Shyamsunder, Kiekintveld Christopher, Ordonez Fernando, and Tambe Milind. ``IRIS-a tool for strategic security allocation in transportation networks." AAMAS (Industry Track) (2009): 37-44, \url{https://doi.org/10.1017/CBO9780511973031.005}.

\bibitem{kiekintveld2009computing}Kiekintveld, Christopher, Jain Manish, Tsai Jason, Pita James, Ord{\'o}nez Fernando, and Tambe Milind. ``Computing optimal randomized resource allocations for massive security games." Proceedings of The 8th International Conference on Autonomous Agents and Multiagent Systems-Volume 1 (2009):689-696.

\bibitem{tsai2010urban}Tsai, Jason, Yin Zhengyu, Kwak Jun-young, Kempe David, Kiekintveld Christopher, and Tambe Milind. ``Urban security: Game-theoretic resource allocation in networked domains." Twenty-Fourth AAAI Conference on Artificial Intelligence (2010).

\bibitem{yang2011improving}Yang, Rong, Kiekintveld Christopher, Ordonez Fernando, Tambe Milind, and John Richard. ``Improving resource allocation strategy against human adversaries in security games." Twenty-Second International Joint Conference on Artificial Intelligence (2011): 458–464.

\bibitem{cappanera2011optimal}Cappanera, Paola, and Scaparra Maria Paola. ``Optimal allocation of protective resources in shortest-path networks." Transportation Science 45, no. 1 (2011): 64-80.

\bibitem{fang2013designing}Fang, Fei, Jiang Albert Xin, and Tambe Milind. ``Designing optimal patrol strategy for protecting moving targets with multiple mobile resources." International workshop on optimisation in multi-agent systems (OPTMAS) (2013).

\bibitem{fang2013protecting}Fang, Fei, Jiang Albert Xin, and Tambe Milind. ``Protecting moving targets with multiple mobile resources." Journal of Artificial Intelligence Research 48, (2013): 583-634.

\bibitem{shang2018operation}Shang, Huilin, Zhang Xiaoke, Ye Ziqing, Zhao Qingsong, Yang Kewei. ``Operation loop-based network design model for defense resource allocation with uncertainty." IEEE Systems Journal 13, no. 1 (2018): 477-488.

\bibitem{yang2012modeling}Yang, Rong, Fang Fei, Jiang Albert Xin, Rajagopal Karthik, Tambe Milind, and Maheswaran Rajiv. ``Modeling human bounded rationality to improve defender strategies in network security games." HAIDM workshop at AAMAS (2012).

\bibitem{dias2014inverted}Dias, Jos{\'e} Capela, Machado Penousal, Silva Daniel Castro, and Abreu Pedro Henriques. ``An inverted ant colony optimization approach to traffic." Engineering Applications of Artificial Intelligence 36, (2014) 122-133.

\bibitem{garcia2012swarm}Garc{\'\i}a-Nieto, Jos{\'e}, Alba Enrique, and Olivera A Carolina. ``Swarm intelligence for traffic light scheduling: Application to real urban areas." Engineering Applications of Artificial Intelligence 25, no. 2 (2012): 274-283.


\end{thebibliography}
\end{document}